\documentclass[a4paper,10pt]{article}
\usepackage[utf8x]{inputenc}

\title{Spontaneous Breaking of Lorentz Symmetry by Ghost Condensation in Perturbative Quantum Gravity}
\author{Mir Faizal\\ Department of Mathematics,  Durham University,\\ Durham, DH1 3LE,  United Kingdom,\\ faizal.mir@durham.ac.uk}

\begin{document}

\maketitle

\begin{abstract}
In this paper we will study the spontaneous breakdown of the Lorentz symmetry by ghost condensation 
in perturbative quantum gravity. Our analysis will be done in the Curci-Ferrari gauge. 
 We will also analyse modification of the BRST and the anti-BRST transformations by the formation of this ghost condensate. It will be shown that even though 
the modified BRST and the modified anti-BRST transformations are not nilpotent, their nilpotency is restored on-shell. 
\end{abstract}



\section{Introduction}
Lorentz invariance is one of the most important symmetries in nature and has been conformed by all experiments. 
However, recently the violation of this symmetry is being intensively discussed \cite{1}. Many new experiments are designed to detect 
any violation of this symmetry \cite{2, 020}.  

The violation of Lorentz symmetry is being studied because it is expected that interactions in string theories 
might lead to the spontaneous breakdown of Lorentz symmetry \cite{1a, 2a}. In fact it is expected that string 
theory will give spacetime a non-commutative structure  and this will in-turn violate Lorentz symmetry \cite{3a}. 
Lorentz symmetry also seems to be violated in many other approaches to quantum gravity \cite{4a}.
In fact the whole program of Horava-Lifshits theory is based on breaking the Lorentz symmetry so that higher order spatial terms can be added to the 
classical Lagrangian density without adding any higher order temporal ones \cite{5b, 6b}.

 The violation of Lorentz symmetry has also been studied in the context
 of gravity coupled to
Chern-Simons term \cite{8}. Initially this theory was initially studied only at an linearized level \cite{9},
 however, recently these results have been generalized to 
include interactions \cite{10}.  Dynamical Lorentz symmetry breaking induced by 
 radiative corrections, in a 
self-interacting fermionic theory has also been studied \cite{6}.  In fact an
 extension of standard model has 
 been constructed where 
through the Higgs mechanism, tensor fields acquire non-zero vacuum expectation values and thus break Lorentz
symmetry spontaneously \cite{7}.  The nature of Nambu-Goldstone bosons 
associated with Lorentz symmetry breaking have also been 
thoroughly studied \cite{1d}-\cite{5d}. Spin-dependent interactions \cite{6d} and spin-independent interactions \cite{2d} induced by
these Nambu-Goldstone bosons have  been investigated. However, so far spontaneous breakdown of the 
 Lorentz symmetry has  not been studied in the context of perturbative quantum gravity.

In this paper we will investigate the spontaneous
 violation of Lorentz symmetry induced by ghost condensation in perturbative quantum gravity. 
Ghost condensation in 
Yang-Mills theories has been studied  in the Curci-Ferrari gauge \cite{11},
 which is obtained  by the inclusion of non-linear terms to the usual Faddeev-Popov Lagrangian density \cite{12}-\cite{16}.
 Ghost condensation has also been studied
 in the context of ghosts associated with higher derivatives, which occur in theories of  modified gravity \cite{17}-\cite{20}. 
However, so far no work has been done on 
 ghost condensation of the Faddeev-Popov ghosts in 
Curci-Ferrari gauge in perturbative quantum gravity.  
The BRST and the anti-BRST 
symmetries for perturbative quantum gravity in linear gauges 
 have been studied by a number of  authors \cite{21}-\cite{23} and their work has been summarized by N. Nakanishi and I. Ojima \cite{24}. 
The BRST and anti-BRST symmetries for perturbative quantum gravity in Curci-Ferrari gauge have
 also been recently studied \cite{23a}.
In this paper we will study  ghost condensation
 and its consequences for perturbative quantum gravity in Curci-Ferrari gauge. 

It may be noted that it is not possible to perform an explicit  violation of Lorentz symmetry as this will be incompatibility with Bianchi identities and
the covariant conservation laws for the energy-momentum and spin-density tensors, whereas spontaneous Lorentz breaking evades this difficulty \cite{01}. 
So there will be no fundamental change in the classical action of this theory. In this way this present work is sightly different from the work on  
Horava-Lifshits gravity which is based on modifying the classical theory. 
\section{ BRST and Anti-BRST Symmetries} 
In this section we will review usual the BRST and the anti-BRST symmetry for perturbative quantum gravity \cite{23a}.  
The Lagrangian density for pure gravity with  cosmological constant $\lambda$ is given by
\begin{equation}
 \mathcal{L} = \sqrt{-g}(R - 2\lambda),
\end{equation}
where we have adopted  units, such that $ 16 \pi G = 1$.
In perturbative gravity one splits the full metric $g_{ab}$ into the metric for the
 background flat spacetime $\eta_{ab}$ and a small perturbation around it being $h_{ab}$. 
The covariant derivatives along with the lowering and raising of indices are compatible with the 
metric for the background spacetime and  small perturbation $h_{ab}$ is viewed as the field that is to be quantized.

All the degrees of freedom in $h_{ab}$ are not physical as the  Lagrangian density for it is invariant under a gauge transformation, 
\begin{eqnarray}
\delta_\Lambda h_{ab} &=&  D^e_{ab}\Lambda_e \nonumber \\ &=& [ \delta^e_b \partial_a  + \delta^e_a \partial _b  
+ \eta^{ce} (\partial_c h_{ab}) + \nonumber \\ &&\eta^{ec} h_{ac}\partial _b  +\eta^{ec} h_{cb} \partial _a ]\Lambda_e, \label{eq}
\end{eqnarray}
where $\Lambda^a$  is a vector field. 
These unphysical degrees of freedom give rise to constraints \cite{25} in the canonical
 quantization  and divergences in the partition function \cite{26} in the path integral quantization.
 So before we can quantize this theory, we need to fix a gauge. This is achieved by  addition of a ghost term and a gauge fixing term to 
the classical Lagrangian density. 
Now let us denote the sum of a ghost term and  a gauge fixing term  by $\mathcal{L}_g$, which is given by 
\begin{eqnarray}
 \mathcal{L}_g &=& -\frac{i}{2}s \overline{s} [h^{ab}h_{ab}]+
 \frac{i\alpha}{2}\overline{s}[b^a c_a]
 \nonumber \\ & =&  \frac{i}{2} \overline{s} s[h^{ab}h_{ab}]- \frac{i\alpha}{2}s[b^a \overline{c}_a], \label{pqr} 
\end{eqnarray}
where  the BRST transformations are given by 
\begin{eqnarray}
s \,h_{ab} &=& D^e_{ab}c_e, \nonumber \\
s \,c^a &=& - c_b \partial ^b c^a, \nonumber \\
s \,\overline{c}^a &=& b^a, \nonumber \\ 
s \,b^a &=&0,
\end{eqnarray}
and the anti-BRST transformations are given by  
\begin{eqnarray}
\overline{s} \,h_{ab} &=&D^e_{ab}\overline{c}_e, \nonumber \\
\overline{s} \,c^a &=& -b^a - 2 \overline{c}_b \partial ^b c^a, \nonumber \\
\overline{s} \,\overline{c}^a &=& - \overline{c}_b \partial ^b \overline{c}^a,\nonumber \\ 
\overline{s} \,b^a &=& - b^b\partial _b c^a.
\end{eqnarray}
In the next section we will analyse   perturbative quantum gravity   in the Curci-Ferrari gauge.
\section{Lorentz Symmetry Breaking}
In order to study spontaneous breaking of Lorentz symmetry, we have to modify the above mentioned BRST and anti-BRST transformations by 
the addition of non-linear terms to them. Thus the  modified BRST transformations are given by
\begin{eqnarray}
s \,h_{ab} &=& D^e_{ab}c_e, \nonumber \\
s \,c^a &=& - c_b \partial ^b c^a, \nonumber \\
s \,\overline{c}^a &=& b^a - \overline{c}^b\partial _b c^a, \nonumber \\ 
s \,b^a &=& - b^b\partial _b c^a -  \overline{c}^b\partial _b c^d\partial _d c^a,
\end{eqnarray}
and the modified anti-BRST transformations are given by
\begin{eqnarray}
\overline{s}\, h_{ab} &=&D^e_{ab}\overline{c}_e, \nonumber \\
\overline{s} \,\overline{c}^a &=& - \overline{c}_b \partial ^b \overline{c}^a, \nonumber \\
\overline{s} \,c^a &=& - b^a - \overline{c}^b\partial _b c^a, \nonumber \\ 
\overline{s} \,b^a &=& - b^b\partial _b \overline{c}^a 
+  c^b\partial _b\overline{c}^d\partial _d  \overline{c}^a.
\end{eqnarray}
Now as we have modified the BRST and the anti-BRST transformations, so the ghost term also gets modified. 
The sum of this modified ghost term and the gauge fixing term, is given by 
\begin{eqnarray}
\mathcal{L}^{(mod)}_g &=&\frac{i}{2}s\overline{s}\left[h^{ab}h_{ab} 
- i \alpha \overline{c}^a c_a \right] \nonumber \\
 &=&-\frac{i}{2}\overline{s} s \left[h^{ab}h_{ab} - i \alpha \overline{c}^a c_a \right].
\end{eqnarray}
It may be noted that just like the Yang-Mills theories in Curci-Ferrari gauge, the perturbative quantum gravity also possess a double BRST symmetry, where the 
gauge fixing term and the modified ghost term is written as a combination of the BRST and the anti-BRST transformations. 
This  Lagrangian density for the sum of the modified ghost and gauge fixing terms in the Curci-Ferrari gauge  is related to the usual 
Lagrangian density for the   ghost and gauge fixing terms as follows,
\begin{eqnarray}
 \mathcal{L}^{(mod)}_g &=&  \mathcal{L}_g + \frac{\alpha}{2} \overline{c}^b\partial_b  c^a. \overline{c}^c\partial_c c_a. 
\end{eqnarray}

Thus apart from the usual Lagrangian density there is a non-linear term in it. However,
we can  linearise this non-linear term by means of Hubbard-Stratonovich 
transformations, as follows
\begin{equation}
 \frac{\alpha}{2} \overline{c}^b\partial_b  c^a. \overline{c}^c\partial_c c_a = - \frac{1}{2 \alpha} \phi^a \phi_a - i \phi^a \overline{c}^b \partial_b c_a.
\end{equation}
The field $\phi^a$ introduced here has a vanishing ghost number and is 
required to be hermitian to maintain the  hermiticity of the total Lagrangian density. 
Thus after using the Hubbard-Stratonovich 
transformations, the Lagrangian density for the sum of the modified ghost and the  gauge fixing terms in the Curci-Ferrari gauge becomes,  
\begin{equation}
 \mathcal{L}^{(mod)}_g = L_{gf} + i \overline{c}^a N_{ab} c^b - \frac{1}{2\alpha}\phi^a \phi_a,
\end{equation}
where 
\begin{eqnarray}
N_{ab} &=& K_{ab} - \phi_a \partial_b,
\end{eqnarray}
here $K_{ab}$ is the contribution coming from the original  ghost term  and is given by 
\begin{equation}
 K_{ab} = \eta_{ae}\eta_{bf}\eta^{nm}\eta^{pq} D^e_{np} D^{f}_{mq}.
\end{equation}

Now we sum over all one-loop ghost diagrams with arbitrary number of external $\phi^a$ fields. This gives us an effective potential $V[\phi]$, which is given by 
\begin{equation}
 \int d^4 x \,V[\phi]  = \int d^4 x \, \frac{1}{2 \alpha}  \phi^a \phi_a + i  \log\left[ \det ( N_{ab}) \right].  \label{ff}
\end{equation}
This effective potential obtained from Eq. $(\ref{ff})$ is divergent and thus has to be regulated. 
The  renormalized effective potential thus obtained, is given by 
\begin{equation}
 V[\phi] = \phi^a \phi_a \left[ \frac{1}{2\alpha} + \frac{1}{32\pi^2}  \left( \log\left( \frac{|\phi|}{4\pi \mu^2}\right) + C \right) \right]
\end{equation} 
The stationary point for this effective potential is given by 
\begin{equation}
 \frac{\delta V[\phi]}{\delta \phi^a } =0.\label{h}
\end{equation}
The Eq. $(\ref{h})$ apart from having the trivial solution $\phi^a =0$, also has the non-trivial solutions $\phi^a = \pm \nu^a$. In semi-classical approximation,
 the field $\phi^a$ is shifted as follows
\begin{equation}
 \phi^a \to \phi_{(cl)}^a + \tilde{\phi}^a,
\end{equation}
 where $\phi_{(cl)}^a$  is the classical field and $\tilde{\phi}^a$ represents the quantum fluctuations to it. 
 The vacuum expectation value of the field $\phi^a$ is required to coincide with the classical field 
so that the vacuum expectation value of the quantum 
fluctuations  vanish. Now in the non-trivial vacuum $\phi^a = \pm \nu^a$, 
 we get a non-vanishing vacuum expecting value for the vector field $\phi^a_{(cl)}$, and this spontaneously breaks the Lorentz symmetry, 
\begin{equation}
\phi^a_{(cl)}  = \langle \phi^a \rangle = \pm \nu^a.
\end{equation}
In this section we showed that  the formation of ghost condensate in perturbative quantum gravity spontaneously break the Lorentz symmetry. 
In the next section we will 
investigate the BRST and anti-BRST symmetries in this phase, where the Lorentz symmetry is spontaneously broken.                                                                                                                                                                                                                                                                                                                                                                                                                                                                                                                                                                                                                                                                              
\section{Modified BRST and anti-BRST Transformations}
The formation of ghost condensate not only spontaneously breaks the Lorentz symmetry but it also spoils the nilpotency of the BRST and the anti-BRST 
transformations. 
However, we will see in this section that the nilpotency of these modified BRST and modified anti-BRST transformations is  restored on-shell. 
The BRST transformations get modified due to the formation of ghost condensates as follows,
\begin{eqnarray}
s \,h_{ab} &=& D^e_{ab}c_e, \nonumber \\
s \,c^a &=& - c_b \partial ^b c^a, \nonumber \\
s \,\overline{c}^a &=& b^a - \overline{c}^b\partial _b c^a, \nonumber \\ 
s \,b^a &=& - b^b\partial _b c^a, \nonumber \\ 
s\, \phi^a &=& 2\psi^a, \nonumber \\ 
s\, \psi^a \, &=& -\frac{1}{2} \psi^b \partial_b \psi^a, \nonumber \\ 
s\,  \overline{\psi}^a &=& \phi^a -i\overline{c}^b \partial_b c^a, 
\end{eqnarray}
and the anti-BRST transformations get modified as follows, 
\begin{eqnarray}
\overline{s}\, h_{ab} &=&D^e_{ab}\overline{c}_e, \nonumber \\
\overline{s} \,\overline{c}^a &=& - \overline{c}_b \partial ^b \overline{c}^a, \nonumber \\
\overline{s} \,c^a &=& - b^a - \overline{c}^b\partial _b c^a, \nonumber \\ 
\overline{s} \,b^a &=& - b^b\partial _b \overline{c}^a,\nonumber \\ 
\overline{s}\, \phi^a &=& 2\overline{\psi}^a, \nonumber \\ 
\overline{s}\, \psi^a \, &=& -\frac{1}{2} \overline{\psi}^b \partial_b \overline{\psi}^a, \nonumber \\ 
\overline{s}\,  \overline{\psi}^a &=& \phi^a -i\overline{c}^b \partial_b c^a. 
\end{eqnarray}
Here $\phi^a$ plays the role of a new Nakanishi-Lautrup field and  and $\psi^a$ and $\overline{\psi}^a$ play the role of new ghosts and anti-ghosts respectively.
These new BRST and anti-BRST transformations are not nilpotent, because 
\begin{eqnarray}
 s^2 \, \overline{\psi}^a &=& 2 \psi^a - b^b \partial_b c^a \neq 0,\nonumber \\ 
\overline{s}^2 \, \psi^a  &=& 2 \overline{\psi}^a + \overline{b}^b \partial_b \overline{c}^a \neq 0. 
\end{eqnarray}
However, their nilpotency is restored by using the equation of motion for these fields  and thus on-shell version of the above two  transformations 
is given by
 \begin{eqnarray}
 \left[s^2 \, \overline{\psi}^a\right]_{on-shell}&=&  0,\nonumber \\ 
 \left[\overline{s}^2 \, \psi^a \right]_{on-shell} &=& 0. 
\end{eqnarray}

The sum of the gauge fixing term and the  ghost term also gets modified because of the modification of the BRST and the anti-BRST transformations.
 However, even after this modification 
the sum  of the gauge fixing term and the  ghost term possess a double BRST symmetry on-shell and
  thus can  be written as a combination of the BRST and the anti-BRST transformations on-shell,
\begin{eqnarray}
 \mathcal{L}^{(new)}_g &=& \frac{i}{2}s\overline{s}\left[h^{ab}h_{ab} - i \alpha \overline{c}^a c_a  -i \frac{\alpha}{2} \phi^a\phi_a  \right] \nonumber \\
 &=&-\frac{i}{2}\overline{s} s \left[h^{ab}h_{ab} - i \alpha \overline{c}^a c_a -i \frac{\alpha}{2} \phi^a\phi_a \right].
\end{eqnarray} 
The sum of the gauge fixing term and modified ghost term is related to the sum of the usual   
 ghost term and gauge fixing term   as follows,
\begin{equation}
  \mathcal{L}^{(new)}_g  = \mathcal{L}_g - i\alpha \phi^a \overline{c}^b \partial_b c_a - \alpha \overline{\psi}^a \psi_a + \frac{\alpha}{2}\phi^a \phi_a.
\end{equation}
It may be noted that the appearance of the term $\alpha \overline{\psi}^a \psi_a$ only produces  a multiplicative overall
factor, which can be absorbed in the normalisation constant of the partition function.   

\section{Conclusion}
We have seen how Lorentz symmetry is spontaneously broken by the formation of ghost condensates in perturbative quantum gravity.
We have also  analysed the  modification of the BRST and the anti-BRST transformations by  this ghost condensation. We have shown that even though 
the modified  BRST and the modified  anti-BRST transformations are not nilpotent, their nilpotency is restored on-shell. 
 
One of the possible signals of Lorentz violation may come from CMB \cite{1e} and other 
high-energy astrophysical observations \cite{2e}.  Violation of  Lorentz might be helpful in explaining 
the polarization of CMB.

 Lorentz violation will also have interesting phenomenological signatures. 
 In fact if there is a  non-vanishing vacuum expecting value for the vector field then  it might lead to a 
decrease in the anomaly frequency of a positron if the anomaly frequency of an electron is increased \cite{3e}. 
However, so far no violation of Lorentz symmetry has been detected  \cite{1f, 2f}.
Another signature of Lorentz symmetry breaking might come from the spectral analysis of  
the spectra of  atoms made up of matter   and similar atoms made up of anti-matter. 
In fact calculations on the spectrum of  hydrogen and anti-hydrogen show that  tiny differences will occur in some lines,
 and no differences will occur in others
if the Lorentz symmetry is spontaneously broken \cite{02}.

The non-vanishing vacuum expecting value for the vector field  could possible explain the occurrence of the cosmological constant. The fact that spontaneous 
breakdown of the Lorentz symmetry  can only occur at very high energies might explain why the cosmological constant has such a small value. 

Our work has been done in flat 
spacetime and it will be interesting to generalise this to general spacetimes or at least maximally symmetric spacetimes like 
the de Sitter spacetime and anti-de Sitter spacetime. A similar analysis might lead to a spontaneous breakdown of the de Sitter or 
anti-de Sitter invariance in those spacetimes.

\end{document}